\documentclass{article}
\usepackage{graphicx}
\usepackage{amsmath}
\usepackage{amssymb}
\graphicspath{{./}{./figs/}}

\def\R{\mathbb{R}}

\def\P{\mathcal{P}}
\def\Q{\mathcal{Q}}
\def\p{p}
\def\q{q}
\def\I{\mathcal{I}}
\def\C{\mathcal{C}}
\def\Dp{\Delta\p}
\def\Dq{\Delta\q}
\def\mybar#1{\boldsymbol{\underline{#1}}}
\def\myhat#1{\widehat{#1}}

\begin{document}
\title{Tracking an Object with Unknown Accelerations
  using a Shadowing Filter} 
\author{Kevin Judd, University of Western Australia}

\maketitle

\begin{abstract}
  A commonly encountered problem is the tracking of a physical object,
  like a maneuvering ship, aircraft, land vehicle, spacecraft or animate
  creature carrying a wireless device. The sensor data is often limited
  and inaccurate observations of range or bearing. This problem is more
  difficult than tracking a ballistic trajectory, because an operative
  affects unknown and arbitrarily changing accelerations. Although
  stochastic methods of filtering or state estimation (Kalman filters and
  particle filters) are widely used, out of vogue variational methods are
  more appropriate in this tracking context, because the objects do not
  typically display any significant random motions at the length and time
  scales of interest. This leads us to propose a rather elegant approach
  based on a \emph{shadowing filter}. The resulting filter is efficient
  (reduces to the solution of linear equations) and robust (uneffected by
  missing data and singular correlations that would cause catastrophic
  failure of Bayesian filters.) The tracking is so robust, that in some
  common situations it actually performs better by ignoring error
  correlations that are so vital to Kalman filters.
\end{abstract}


\section{Introduction}
The literature on tracking of isolated or multiple objects in uncluttered
and cluttered environments is broad and varied, with methodological
approaches ranging from variational techniques to an extensive variety of
Kalman filters, particles filters, and other sequential Bayesian
filters~\cite{Anderson-Moore:optimal-filtering, BarSalom-etal:tracking,
  Jazwinski:filtering, Mahler:information-fusion}.

Optimal tracking (filtering or state estimation) requires identification
of the appropriate length and time scales of a object's motion, and the
magnitudes of the uncertainty (observational and dynamical noise) and the
nonlinearity at these scales. There are three broad scenarios: (1)~when
the update cycle is fast, such in guidance control, Kalman filters are
appropriate, because system behaviours are almost linear; (2)~when
nonlinearity is significant, and the noise largely dynamical, as opposed
to observational, then particle filters are appropriate; and (3)~when
nonlinearity is significant, but the noise largely observational, then
variational methods are most appropriate.
Mathematically situations (1) and~(2) best assume the underlying process
is stochastic, whereas situation~(3) it is best to assume a deterministic
dynamical system: the success and universal acceptance of stochastic
methods, over the past decades, has lead to them being applied in
situation~(3), where they are not the most appropriate
choice~\cite{Judd-Stemler:deadparrot2}.

Here we present an approach to tracking in situation~(3) based on
\emph{shadowing filters}, which derive from the modern concept of
shadowing in dynamical systems theory; literally meaning \emph{to find a
  trajectory that \textbf{shadows} the
  observations}~\cite{Gilmour:PhD}. The methodology has its roots in the
work of Laplace and Gauss fitting celestial orbits as
curves~\cite{Davis:Gauss}, and subsequent least squares approaches for
ballistic trajectories. 
Shadowing filters lie within the domain of variational methods, as in
optimal control, 
but are subtly different~\cite{Judd:dcip}. Shadowing filters are not
equivalent to 4D-variational assimilation often employed in meteological
and oceanographic modelling and forecasting. 
Nor are they equivalent to dynamic programming approaches that finds a
Viterbi path~\cite{Forney:Viterbi-algorithm}.

To develop our methodology, a one-dimensional, or scalar case, is
considered first, which is then extended to the multi-dimensional vector
case, where the observations are of the components of the Cartesian
position vector. From this basis other relevant observation situations
that are often encountered are considered, for example, using range or
bearing observations from one or more sensors.  A significant problem that
arises in these types of observation networks is the way the covariance of
observational errors varies with position, in particular, the
singularities that occur when the target and sensors are co-linear or
co-planar.  Possibly the most surprising outcome of the shadowing filter
approach is that singular covariances of observations are not a
significant problem, and targets can be tracked through missing data and
singularities relatively easily; indeed, sometimes covariances can be
ignored entirely, obtaining very efficient tracking filters.

To keep the exposition of the algorithm and its benefits clear, we
restrict attention to tracking an isolated vehicle in an uncluttered
environment. It should be clear, once the methodology is understood, that
since the shadowing filter assumes a tracked object maintains a contiguous
trajectory it will perform well with multiple targets in cluttered
environments.

\section{Formulation and implementation}

\subsection{Scalar case}

Our initial interest is tracking the position of a point object in one
dimension, given a sequence of noisy observations. Let $\P_i\in\R$ be the
observed position at time~$t_i$ for $i=0,\dots,n$, and $\sigma_i^2$ be the
variance of the observational error. The object's dynamics are modelled by
its position $\p_i\in\R$ and velocity $v_i\in\R$ at $t_i$, and constant
acceleration $a_i\in\R$ for $t_i\leq{}t<t_{i+i}$. For notational
convenience, define $\tau_i=t_{i+1}-t_i$. Our goal is to have $\p_i$ close
to $\P_i$ subject to the accelerations not being excessively large or
changeable. We might therefore choose to minimise the total square error
$\sum_{i=0}^n\sigma_i^{-2}(\P_i-\p_i)^2$ subject to accelerations over the
interval being bounded, $a_i^2\leq\xi^2$ $i=0,\dots,n-1$. Bounded
acceleration is an appropriate constraint, but it introduces technical
difficulties. Although these difficulties can be
over-come~\cite{Judd:dcip}, it is more convenient and efficient to instead
constrain the root mean squared acceleration over the entire trajectory,
$\sum_{i=0}^{n-1}\tau_ia_i^2\leq{}(t_n-t_0)\xi^2$.

Assuming Newton's laws and Galilean transforms apply to the point object's
motion, then the stated optimisation problem can be posed using a
Lagrangian:
\begin{align}
\label{eq:L}
L
&=
	\frac{1}{2}\sum_{i=0}^n\sigma_i^{-2}(\P_i-\p_i)^2 \\
	& + \sum_{i=0}^{n-1} \lambda_{i}(\p_{i+1}-\p_{i}-v_{i}\tau_{i}-\frac12 a_{i}\tau_{i}^2) \\
	& + \sum_{i=0}^{n-1} \mu_{i}(v_{i+1}-v_{i}-a_{i}\tau_{i}) \\
	& + \eta \left(\sum_{i=0}^{n-1}\tau_ia_i^2-(t_n-t_0)\xi^2\right), \label{eq:L4}
\end{align}
where $\lambda_i\in\R$, $\mu_i\in\R$ and $\eta\geq0$ are dual
variables. Solving the optimisation, and defining
\begin{equation}
  \label{eq:spline}
  p(t) = p_i+v_i(t-t_i)+\frac12a_i(t-t_i)^2 \quad\text{for}\quad t_i\leq t\leq t_{i+1},
\end{equation}
provides an optimal quadratic spline estimate of a particle's path
assuming piecewise constant accelerations.
The optimal solution occurs where all the partial derivatives of $L$ are
zero: \newcommand{\pd}[2]{\frac{\partial#1}{\partial#2}}
\begin{align}
\label{eq:Lx}
\pd{L}{\p_i} &= 
\left.\begin{cases}
-\sigma_0^{-2}(\P_0-\p_0) - \lambda_0, & i=0,\\
-\sigma_i^{-2}(\P_i-\p_i) + \lambda_{i-1} - \lambda_{i}, & 0<i<n,\\
-\sigma_n^{-2}(\P_n-\p_n) + \lambda_{n-1}, & i=n,\\
\end{cases} \right\}=0 \\
\label{eq:Lu}
\pd{L}{v_i} &= 
\left.\begin{cases}
-\lambda_{0}\tau_{0}-\mu_{0}, & i=0,\\
-\lambda_{i}\tau_{i}+\mu_{i-1}-\mu_{i}, & 0<i<n,
\end{cases}\right\}=0 \\
\label{eq:La}
\pd{L}{a_i} &=
-\frac12\lambda_{i}\tau_{i}^2 - \mu_{i}\tau_{i} + 2\eta{}\tau_ia_{i}=0,\\
\label{eq:Ll}
\pd{L}{\lambda_i} &= \p_{i+1}-\p_{i}-v_{i}\tau_{i}-\frac12 a_{i}\tau_{i}^2=0,\\
\label{eq:Lm}
\pd{L}{\mu_i} &= v_{i+1}-v_{i}-a_{i}\tau_{i}=0,\\
\label{eq:Le}
\pd{L}{\eta} &= \sum_{i=0}^{n-1}\tau_ia_{i}^2-(t_n-t_0)\xi^2=0.
\end{align}
Equation~(\ref{eq:Lx}) is defined for $0\leq{}i\leq{}n$ while
(\ref{eq:Lu}--\ref{eq:Lm}) are defined for $0\leq{}i<n-1$. With the
exception of~(\ref{eq:Le}), the remaining five equations are linear in the
unknowns. However, $\xi$ and $\eta$ are related through term~(\ref{eq:L4})
of the Lagrangian, which is the only place they appear. Hence, one can
solve the linear equations~(\ref{eq:Lx}--\ref{eq:Lm}) for a fixed $\eta$,
then compute the corresponding value of $\xi$ from eq.~(\ref{eq:Le}). The
optimal solution for any $\xi$ can be approximated arbitrarily closely by
an efficient one-dimensional search, such as Brent's
method~\cite{Press-etal:numerical-recipes}. In practice it is unlikely
that $\xi$ needs to be specified precisely, after all, it is only a bound
on the root mean squared acceleration. Consequently, it is usually
sufficient to work only with $\eta$, treating it as a smoothing, or
regularisation, parameter.

Combining (\ref{eq:Ll}) and (\ref{eq:Lm}) to eliminate the $v_i$
gives\footnote{To do this multiply (\ref{eq:Ll}) by $\tau_{i-1}$, then
  take another copy of (\ref{eq:Ll}) with $i$ replaced with $i-1$,
  multiply by $\tau_i$, and substract this from the former. Then use
  (\ref{eq:Lm}) to eliminate $v_i-v_{i-1}$.}
\begin{equation}
	\label{eq:xa}
	\p_{i+1}\tau_{i-1}-\p_i(\tau_i+\tau_{i-1})+\p_{i-1}\tau_i 
    = \frac12(a_i\tau_i+a_{i-1}\tau_{i-1})\tau_{i-1}\tau_i,
\end{equation}
for $0<i<n$.  Combining (\ref{eq:Lx}), (\ref{eq:Lu}) and (\ref{eq:La}) and
eliminating the dual variables $\lambda_i$ and $\mu_i$ (as explained in
the following), obtains another set of expressions relating the $\p_i$ and
$a_i$, which when combined with~(\ref{eq:xa}) enable solving for a near
optimal solution very efficiently.

There is a certain amount of redunancy in the equations just stated, but
to assist formulation of a solution using matrix notation it is
advantageous to retain the redundancy.

Define column vectors $\P=(\P_0,\dots,\P_n)^T\in\R^{n+1}$,
$\p=(\p_0,\dots,\p_n)^T\in\R^{n+1}$,
$\lambda=(\lambda_0,\dots,\lambda_{n-1})^T\in\R^n$,
$\mu=(\mu_0,\dots,\mu_{n-1})^T\in\R^n$, and finally
$a=(a_0,\dots,a_{n-1})^T\in\R^{n}$. Define $\tau$ to be the
$n\times{}n$ matrix of zeros with main diagonal
$(\tau_0,\dots,\tau_{n-1})$, and $\I$ to be the $(n+1)\times(n+1)$ matrix
of zeros with main diagonal $(\sigma_0^{-2},\dots,\sigma_{n}^{-2})$. Define a
$n\times{}n$ matrix $D$ to have all entries zero except $-1$ on the main
diagonal and $1$ on the first lower diagonal, and similarly define a
$(n+1)\times{}n$ matrix $E$: specifically
\begin{equation}\label{eq:DE}
D_{ij} = E_{ij} = 
\begin{cases}
-1, & i=j,\\
 \phantom{-}1, & i=j+1,\\
 \phantom{-}0, & \text{otherwise,}
\end{cases}
\end{equation}
when the entry is defined.  It follows that the linear equations
(\ref{eq:Lx}), (\ref{eq:Lu}) and~(\ref{eq:La}) can be succinctly expressed
as
\begin{equation}
\label{eq:succinct}
E\lambda = \I(\P-\p), \qquad D\mu=\tau\lambda,
\qquad 2\eta\tau{}a=\frac12\tau^2\lambda+\tau\mu.
\end{equation}
In the last of the three equations of~(\ref{eq:succinct}), $\tau$ is
invertible, so that a $\tau$ factor can be canceled on the left of each term.

Define a $n\times{}n$ matrix $L$, and $n\times{}(n+1)$ matrix $M$, to have
a lower triangular form:
\begin{equation}\label{eq:ML}
L_{ij} = M_{ij} = 
\begin{cases}
 -1, & i\geq{}j,\\
 \phantom{-}0, & \text{otherwise,}
\end{cases}
\end{equation}
when the entry is defined. It can be easily verified that $DL=I$ and
$EM=J$ where $I$ is the $n\times{}n$ identity matrix, and $J$ is a
$(n+1)\times(n+1)$ matrix with
\begin{equation}\label{eq:J}
J_{ij} = 
\begin{cases}
 \phantom{-}1, & i=j \text{ and } i\neq{}n+1,\\
 -1, & i=n+1 \text{ and } j\neq{}n+1,\\ 
 \phantom{-}0, & \text{otherwise.}
\end{cases}
\end{equation}
The identities $DL=I$ and $D\mu=\tau\lambda$ imply\footnote{If $DL=I$,
  then $DL\tau\lambda=\tau\lambda$, but $D\mu=\tau\lambda$, implying
  $\mu=L\tau\lambda$.}  $\mu=L\tau\lambda$. The identities $EM=J$ and
$E\lambda=\I(\P-\p)$ imply\footnote{If $EM=J$, then
  $EM\I(\P-p)=J\I(\P-p)$. If $E\lambda=\I(\P-p)$ also, then, since the
  first $n$ rows of $J$ is the $n\times{}n$ identity,
  $\lambda=M\I(\P-p)$. Substituting this $\lambda$ back into
  $E\lambda=\I(\P-p)$, gives $J\I(\P-p)=\I(\P-p)$; the first $n$ rows are
  a tautology, but the last implies
  $\sum_{i=0}^{n-1}\sigma_i^{-2}(\P_i-\p_i)=0$.} that $\lambda=M\I(\P-\p)$
and $\sum_{i=0}^{n-1}\sigma_i^{-2}(\P_i-\p_i)=0$.
It follows, by substitution into
the last equation of~(\ref{eq:succinct}), that
\begin{equation}
\label{eq:a2}
2\eta{}a = \left(\frac12\tau{}M+L\tau{}M\right)\I(\P-\p),
\end{equation}
which can be combined with (\ref{eq:xa}) as follows.
Define $(n-1)\times{}(n+1)$ matrices $A$ and $B$, and $(n-1)\times{}n$
matrix $G$:
\begin{equation}\label{eq:G}
G_{ij} = 
\begin{cases}
 \tau_{i}^2\tau_{i+1}, & i=j,\\
 \tau_{i}\tau_{i+1}^2, & i+1=j,\\
 \phantom{}0, & \text{otherwise,}
\end{cases}
\end{equation}
\begin{equation}\label{eq:B}
  B_{ij} = 
\begin{cases}
 \tau_{i+1}, & i=j,\\
 -(\tau_i+\tau_{i+1}), & i+1=j,\\
 \tau_{i}, & i+2=j,\\
 0, & \text{otherwise,}
\end{cases}
\end{equation}
\begin{equation}\label{eq:A}
A = \frac14 G\left(\frac12\tau{}M+L\tau{}M\right).
\end{equation}
Then (\ref{eq:xa}) can be written $B\p=\frac12Ga$, which when combined
with (\ref{eq:a2}) obtains the equation
\begin{equation}
  \label{eq:semi-master}
  (A\I+\eta{}B)\p = A\I\P,
\end{equation}
and the additional constraint
$\sum_{i=0}^{n-1}\sigma_i^{-2}(\P_i-\p_i)=0$.  For stability reasons
discussed in section~\ref{sec:presentation}, this constraint will be
extended to $\sum_{i=0}^{n}\sigma_i^{-2}(\P_i-\p_i)=0$. Defining a
$n\times{}(n+1)$ matrix $\mybar{B}$ to be $B$ augmented with a final row
of zeros, and a $n\times{}(n+1)$ matrix $\mybar{A}$ to be $A$ augmented
with a final row 
of ones, then
\begin{equation}
  \label{eq:master}
  \left(\mybar{A}\I+\eta\mybar{B}\right) \p = \mybar{A}\I\P,
\end{equation}
encodes both (\ref{eq:semi-master}) and the extended
constraint. Solving~(\ref{eq:master}) by singular value decomposition
obtains a least squares approximate solution for the position~$\p$ for
given smoothing parameter~$\eta$. See section~\ref{sec:presentation} on
the nature of this approximation and the optimal presentation of the data
in~$\P$. \emph{It is \textbf{important} to read
  section~\ref{sec:presentation} before implementing~(\ref{eq:master}).}

\subsection{Vector case}

Consider now the situation where a point vehicle is positioned in a
$d$-dimensional Euclidean space with Cartesian coordinates. Suppose that
the coordinate positions are observed as a sequence $\P_i\in\R^d$ in such
a way that the observational errors have a $d\times{}d$ covariance matrix
$\C_i$, and corresponding information matrix $\I_i=\C_i^{-1}$. The
quantities to be determined are $p_i\in\R^d$, $v_i\in\R^d$, and
$a_i\in\R^d$, which are all now $d$-dimensional column vectors. If the aim
is to track the trajectory under the assumption of bounded RMS magnitude
of the acceleration, the Lagrangian~(\ref{eq:L}) now becomes vectorised as
\begin{eqnarray}
\label{eq:Ld}
L
&=& 
	\frac{1}{2}\sum_{i=0}^n(\P_i-\p_i)^T\I_i(\P_i-\p_i) \\
	&& + \sum_{i=0}^{n-1} \lambda_{i+1}^T(\p_{i+1}-\p_{i}-v_{i}\tau_{i}-\frac12 a_{i}\tau_{i}^2) \\
	&& + \sum_{i=0}^{n-1} \mu_{i+1}^T(v_{i+1}-v_{i}-a_{i}\tau_{i}) \\
	&& + \eta \left(\sum_{i=0}^{n-1}\tau_ia_i^Ta_i-(t_n-t_0)\xi^2\right), \label{eq:Ld4}
\end{eqnarray}
where $\lambda_i\in\R^d$, $\mu_i\in\R^d$ are now $d$-dimensional column
vectors, but $\eta\in\R$. The superscript $T$ indicates the transpose.

The solution of the vectorised optimization problem proceeds identically
to the scalar case, using the same linear algebra methods. Let
$\P\in\R^{d(n+1)}$ denote a column vector being the time-series of
$n+1$~observations~$\P_i\in\R^d$ stacked in $d$-dimensional blocks, and
similarly for position variables~$p\in\R^{d(n+1)}$. Let $\I$ denote the
$(n+1)d\times(n+1)d$ block diagonal matrix with the $d\times{}d$
information matrices $\I_i$ along the diagonal. Finally, let $I_d$
denotes the $d\times{}d$ identity matrix and let $\myhat{M}=M\otimes{}I_d$
denote the \emph{outer product} of an arbitrary matrix~$M$ with~$I_d$,
that is, $\myhat{M}$ has a block structure where each scalar
entry~$M_{ij}$ of~$M$ becomes a $d\times{}d$-block $M_{ij}I_d$ of
$\myhat{M}$. Then the vectorised solution is
\begin{equation}
  \label{eq:masterd}
  \left(\mybar{\myhat{A}}\I+\eta\mybar{\myhat{B}}\right) \p = \mybar{\myhat{A}}\I\P.
\end{equation}

\subsection{Implementation and interpretation of the filter}\label{sec:presentation}

Solving (\ref{eq:master}) or~(\ref{eq:masterd}) obtains an
\emph{approximate} solution to the optimal shadowing trajectory for a
given smoothness~$\eta$. To see this, note that the system of
equations~(\ref{eq:master}) is under-determined: there are $n$~linear
equations in $n+1$ unknowns. This occurs because in deriving~(\ref{eq:xa})
the velocity variables were eliminated, but to completely define a
trajectory the velocity needs to be known at some time; usually the
initial or final velocity is specified or solved for. To solve for the
velocity requires introducing another $n$~variables and $n$~equations to
solve for all the velocities, which is significant additional computation
for very little benefit. The approximation~(\ref{eq:master}) relies on the
fact that if the time window of the trajectory is sufficiently long, then
accurate specification of the initial velocity is not required. It just
means the initial part of the trajectory may not accurately fit the
observations. However, leaving the initial velocity unspecified can lead
to instability for short time windows. Imposing the extended constraint
$\sum_{i=0}^{n}\sigma_i^{-2}(\P_i-\p_i)=0$ overcomes possible instability
by implicitly defining an initial velocity.

In formulating the Lagrangian forward differences were used to express
position in terms of velocity and acceleration.  Unfortunately, this
results in $A$ and~$B$ having a lower triangular form. Consequently, the
approximation errors are largest for the $\p_i$ with largest $i$, which is
not what is wanted for state estimation and forecasting; it is preferable
that the smallest errors are at the most recent times.  Reformulating the
Langrangian with backward differences solves this problem, however, there
is much simpler solution: initially reversing the time-series data
sequence $(\tau_i,\P_i,\I_i)$, applying the filter (\ref{eq:master}) or
(\ref{eq:masterd}), then reversing shadowing trajectory time-series
$\p_i$ to obtain the desired result. Even this trick is unnecessary.
Let $R$ denote the matrix that reverses a vector, then the time-series
reversal trick, is equivalent to changing (\ref{eq:semi-master}) to
\begin{equation}
  \left(A\I+\eta{}B\right)R\p = AR\I\P,
\end{equation}
but since $R^{-1}=R$, multipling on the left by~$R$ obtains
\begin{equation}
 \left(RAR\I+\eta{}RBR\right)\p = RAR\I\P,
\end{equation}
where the matrices $RAR$ and~$RBR$ are just $A$ and $B$ with their rows
and columns reversed. Hence, the time-reversal trick is some bookkeeping
when constructing $A$ and~$B$.

\section{Illustrative examples}

This section provides demonstrations of the use of the proposed
methods. The scalar filter is considered first, both as an off-line
smoothing filter and sequential state-estimator. The vector filter is
considered for observations in cartesian coordinates, with and without
correlation, which is a preliminary to section~\ref{sec:noncartesian}
where non-cartesian observations are considered.

\subsection{Scalar filter for moothing and sequential tracking}

Here tracking of one observed variable is examined for increasing values
of smoothing paramter~$\eta$. The filter is employed as smoothing filter
over the entire observation window, and as a sequential
state-estimator. In both cases the time-reversal trick discussed in
section~\ref{sec:presentation} is employed. In this example employs a
large red noise component~$\chi_t$ to mimic a vehicle maneuvering in an
unpredictable way.

\begin{figure}
  \centering
  \includegraphics[width=\linewidth]{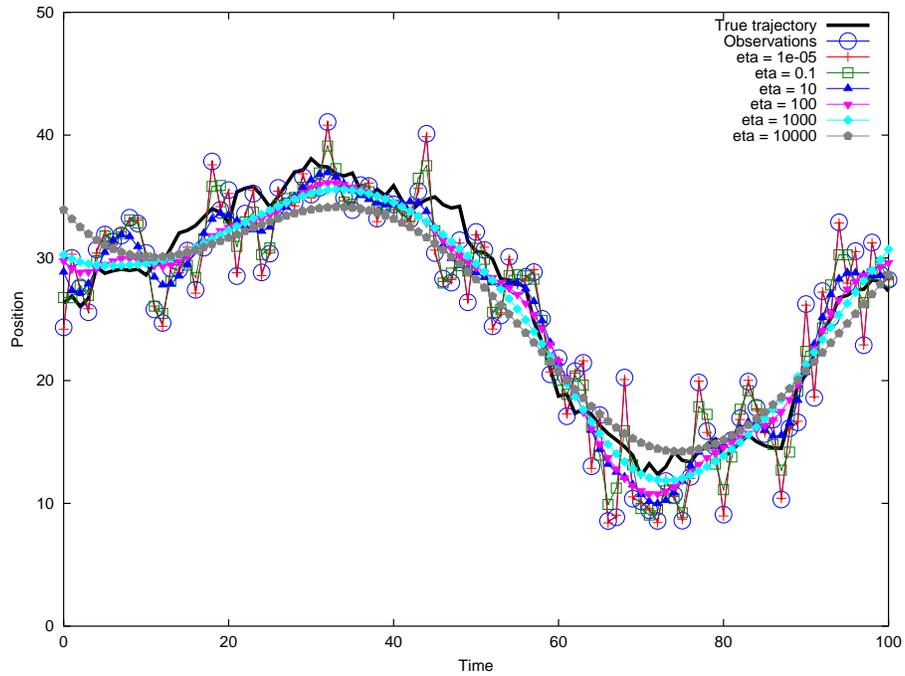}
  \caption{Position tracking $25+10\sin(t/15)+\chi_t+3\epsilon_t$,
    $0\leq{}t\leq{}100$ where $\epsilon_t$ is a white noise process
    $N(0,1)$, and $\chi_t$ an independent cumulative of a white noise
    process $N(0,1)$.  Shadowing filter results for smoothing parameters
    as stated.}
  \label{fig:1a}
\end{figure}

\begin{figure}
  \centering
  \includegraphics[width=\linewidth,trim=0 120 0 120]{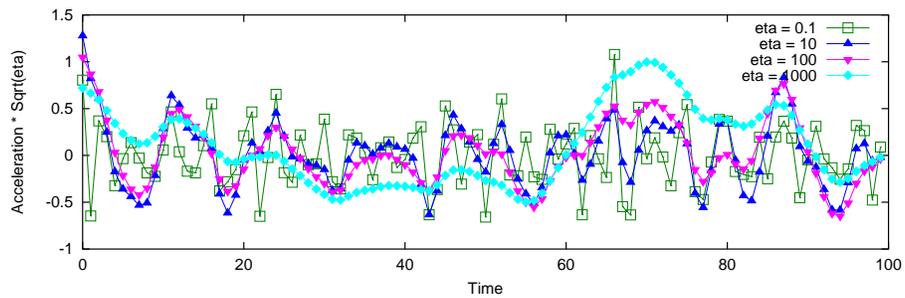}
  \caption{Computed accelerations for tracking shown in fig.~\ref{fig:1a}
    for selected~$\eta$. Accelerations are scaled by $\sqrt{\eta}$ to
    allow easier comparision.}
  \label{fig:1b}
\end{figure}

\begin{figure}
  \centering
  \includegraphics[width=\linewidth,trim=30 20 220 165]{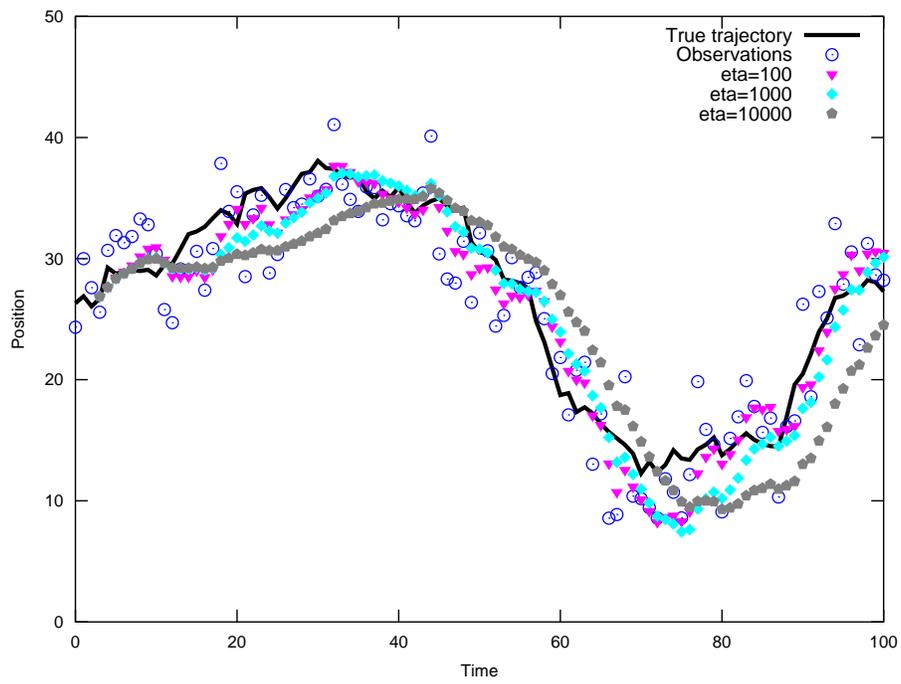}
  \caption{Sequential tracking of same data as shown in fig.~\ref{fig:1a}
    for selected~$\eta$. State estimates use only observations up to that
    time, and final filtered state at that time is plotted.}
  \label{fig:1c}
\end{figure}

Figure~\ref{fig:1a} reveals how the implied approximation in the
solution~(\ref{eq:master}) results in the position tracking deviating from
ideal at the beginning ($t=0$) of the time series. Without the
time-reversal trick, this deviation would have occurred at the end
($t=n$); although the deviation is small, it is significant, but if the
time-series window is large enough, then there is no significant effect
for $t>10$. Optimal smoothing appears occur in the range $10<\eta<100$.

Figure~\ref{fig:1b} shows how smaller~$\eta$ result in large and
rapidly switching accelerations, while larger~$\eta$ result in much smaller
accelerations applied over longer periods.

Figure~\ref{fig:1c} demonstrates using the same filter as in
figure~\ref{fig:1a} as a sequential state-estimator; the filter is
applied only to the observations up to that time. In the this tracking
mode the smoothing parameter~$\eta$ is seen to act like an inertial
damping. For the larger $\eta=10000$ the tracking lags the true
trajectory. For smaller $\eta$ values the tracking is better, but note how
a sequence of observations with repeated negative bias for $60<t<75$
result in the tracking over-shotting the turn near $t=70$. When the
repeated bias ends around $t=75$ the near optimal $\eta=100$ track jumps
back to good estimates, while the $\eta=1000$ track turns back smoothly
toward the true trajectory.

\subsection{Vector filter with uncorrelated observations}
\label{sec:indept}
If the observations of each component of the $d$-dimensional position are
uncorrelated, then filtering can be accomplished very efficiently using a
scale filter, which will come in useful later when non-cartesian
observations are considered. 

When the observations of each component of the $d$-dimensional position
are uncorrelated the covariance matrices $C_i$ are all diagonal, and it is
unnecessary to use the vectorised filter~(\ref{eq:masterd}), which has
been expanded by an outer product with~$I_d$; instead it is sufficient to
solve ~(\ref{eq:master}) seperately for each component. If all the
components have proportionally the same variance at each time, then the
singular value decomposition only needs to be solved once for all
components, that is, one information matrix~$\I$ is needed, whose elements
are proportional to the variances, and $\P$ becomes a $(n+1)\times{}d$
matrix of observations, so that eq.~(\ref{eq:master}) then solves for all
components of~$\p$ simultaneously.

\begin{figure}
  \centering
  \begin{tabular}{ll}
    (a) & (b) \\
    \includegraphics[trim={130 0 110 0},width=0.4\linewidth]{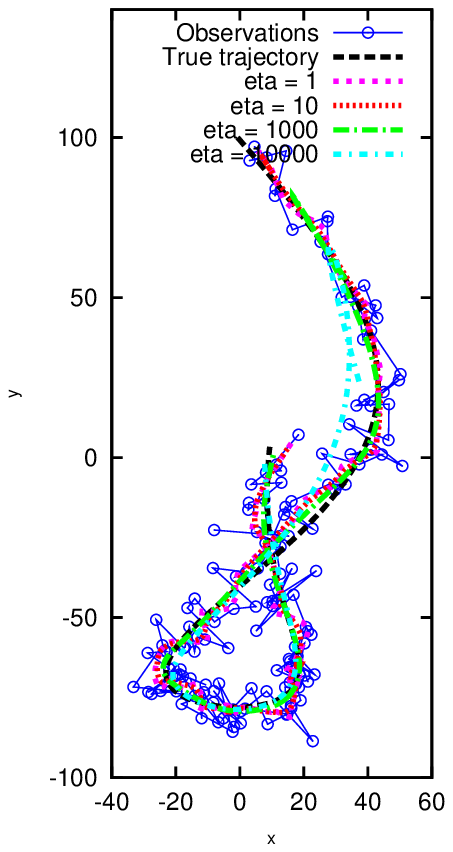}
    &
    \includegraphics[trim={130 0 110 0},width=0.4\linewidth]{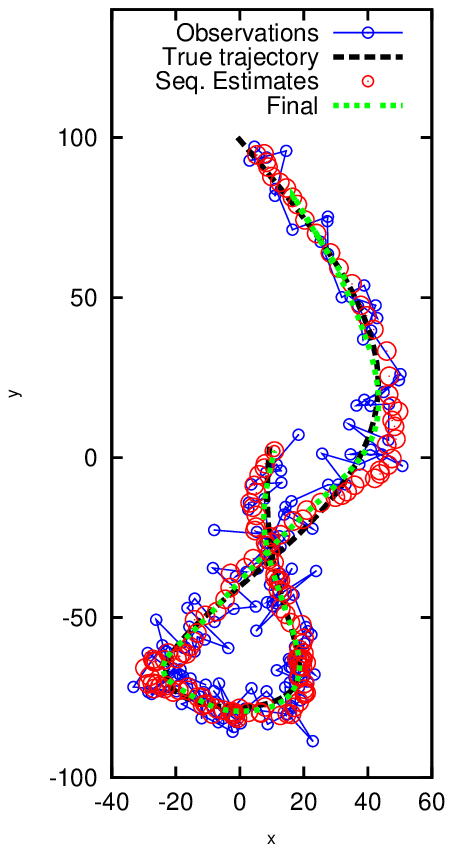}
  \end{tabular}
  \caption{Position tracking for $0\leq{}t\leq{}150$ of the path
    $(x,y)=10(t-10)/150+(1/3)(1-t) (\sin(t/15),2-t/15).$
    Observational errors independent on each component with standard
    deviation~$5$. (a)~Final tracking curves for various
    $\eta$. (b)~Sequential position estimates for $\eta=1000$, that is,
    last state of a shadowing trajectory obtained using all observations
    up to a given time.}
  \label{fig:p2d}
\end{figure}

Figure~\ref{fig:p2d} shows tracking in two dimensions in this
situation. Observe how a very large $\eta=10000$ results in poor tracking
as the tracking curve is pulled toward the mean of the observations. The
poor tracking at the beginning can also be observed for
larger~$\eta$. Fig~\ref{fig:p2d}(b) shows the important case of sequential
estimates, that is, sequential tracking of the object using all the
observations obtained up to a given time. For efficiently reasons, one
would in practice only use a finite window of past observations. Details
of how to determine the optimal window for a given system and purpose is
beyond the scope of this paper and is discussed in the general context of
shadowing filters elsewhere~\cite{Stemler-Judd:guide1}.

\section{Using non-Cartesian observations}\label{sec:noncartesian}

A frequently encountered tracking problem involves using bearing and range
observations, or combinations of multiple bearing or range
observations. Several important situations are worth considering. Active
radar location uses range and bearing information. Satellite
interferometry uses range and bearing information, but the range is much
more accurately measured than the bearing. Global positioning by satellite
uses only range information, but from multiple reference
satellites. Tracking wireless devices can use range information inferred
from signal power at multiple transponders.  Passive sonar location
provides bearing information, but poor range; often bearings from multiple
sensor points are used. 

All of the applications mentioned can be dealt with using the vector
filter (\ref{eq:masterd}), by transforming the observations into \emph{raw
  Cartesian position estimates} and computing the appropriate information
matrix. The transformations are simple geometry, but computing the
information matrices requires some approximation or restrictions.

Let $\p\in\R^d$ be the position in Cartesian coordinates, and let
$\q\in\R^d$ be a vector of $d$ noise-free observations of the position is
some other coordinates. Suppose there is an invertible function~$f$, on
some domain, such that $\p=f(\q)$. Given a noisy observation
$\Q=\q+\Delta\q$, the transform~$f$ provides a raw position
estimate~$\P=f(\Q)=\p+\Delta\p$. Given the covariance matrix~$C_q$ of the
observations, the covariance $C_p$ of the estimate is required. The column
vector $\Dq$ is the error in the observation, and to a first
approximation, the error in the estimate is $\Dp\approx{}J\Dq$, where
$J=\partial_qf(q)$ is the Jacobian matrix of~$f$ at~$\q$.  Since the
covariance $C_p$ is the expected value of the outer product $\Dp\Dp^T$, it
follows that, to a first approximation, $C_p=JC_qJ^T$. It also follows that
the corresponding information matrices are related by $\I_p=K^T\I_qK$,
where $K=J^{-1}=\partial_p(f^{-1})$.
Note that since only the information matrix is needed in the shadowing
filter under discussion, it is sometimes easier to compute the $K$
directly using $f^{-1}$, than it is to compute $J$ and invert it.  This is
the case in some of the following examples.

An important problem, which will be returned to in each of the following
sections, is that although the correlation matrix~$C_q$ may be well known,
the transformation matrices $J$ and~$K$ depend on the target's location,
which is unknown. The raw position estimate $\P=f(Q)$ could be used, but
this introduces errors in the supposed covariance. When the transformed
coordinates are highly correlated and the transform very non-linear, then
a small error in the raw position estimate can give rise to a very wrong
estimate in the correlation. This problem plagues Kalman filters, and
other filters that need a covariance estimate to reliably estimate the
state. A significant advantage of a shadowing filter is the robustness
gained from finding a shadowing trajectory, rather than just a current
position estimate. This robustness means that the correlation in the raw
position estimates are of little importance, that is, ignoring the
correlation can have little effect on the quality of the tracking.

\subsection{Range and bearing observations}
Consider a target tracked in the plane, position $\p=(x,y)$, using
observations $q=(r,\theta)$, where $r$ is the range from a reference point
$(a,b)$ and $\theta$ the bearing in radians measured in the anti-clockwise
direction from the $x$-axis.  The transformation $p=f(q)$ is given by
\begin{eqnarray}
  x &=& a + r\cos\theta,\\
  y &=& b + r\sin\theta.
\end{eqnarray}
Under the assumption that $r$ is not close to zero, and the variances of
$r$ and $\theta$ are small, then the covariance and information matrix of
$x$ and~$y$ are approximated as previously described using
\begin{equation}
  J = 
  \begin{pmatrix}
    \cos\theta & -r\sin\theta\\
    \sin\theta & r\cos\theta
  \end{pmatrix}
\end{equation}
or
\begin{equation}
  K = 
  \begin{pmatrix}
    \cos\theta & \sin\theta\\
    -(1/r)\sin\theta & (1/r)\cos\theta
  \end{pmatrix}.
\end{equation}

\begin{figure}
  \centering
  \begin{tabular}{lll}
    (a) & (b) & (c) \\
    \includegraphics[trim={220 30 200 30},width=0.3\linewidth]{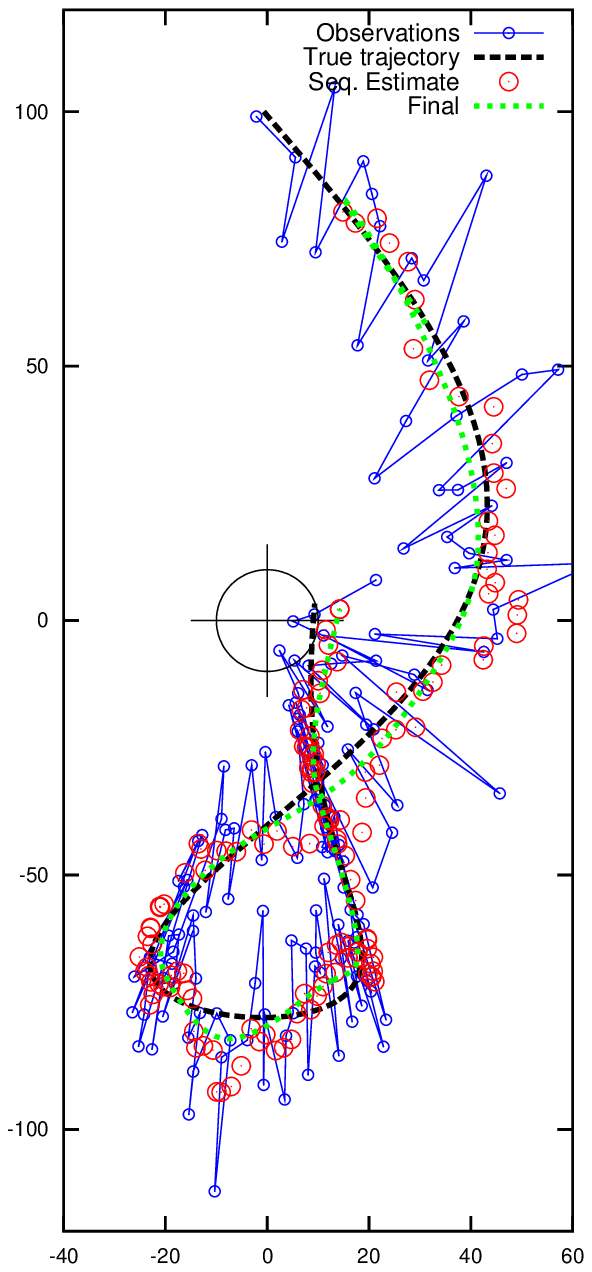}
    &
    \includegraphics[trim={220 30 200 30},width=0.3\linewidth]{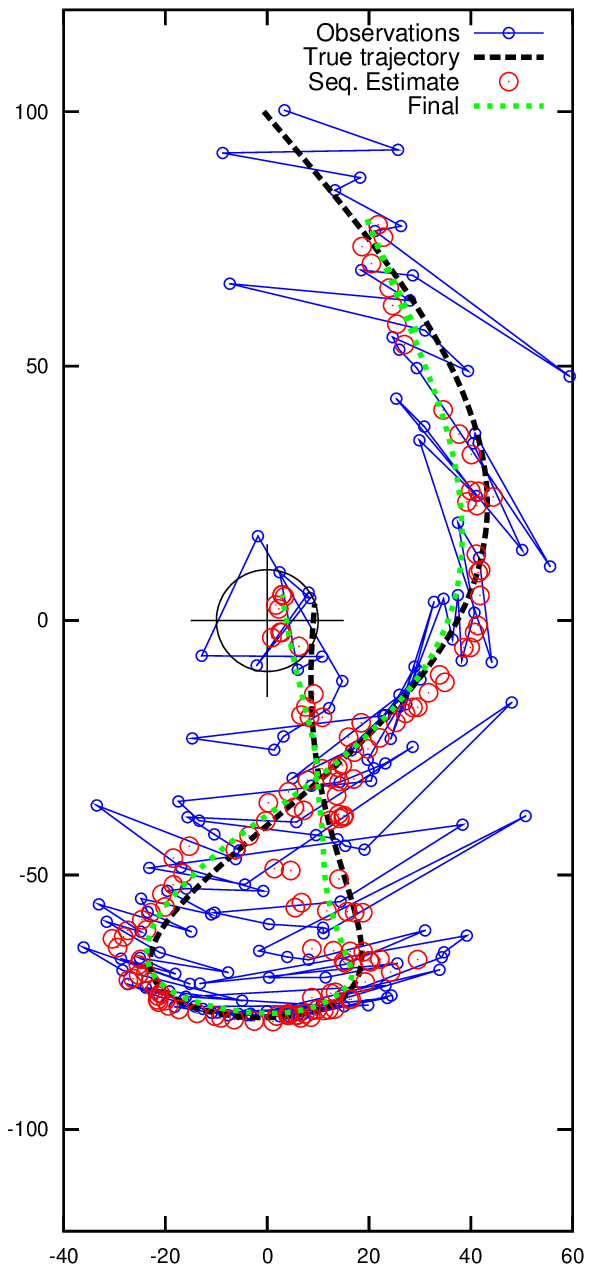}
    &
    \includegraphics[trim={220 30 200 30},width=0.3\linewidth]{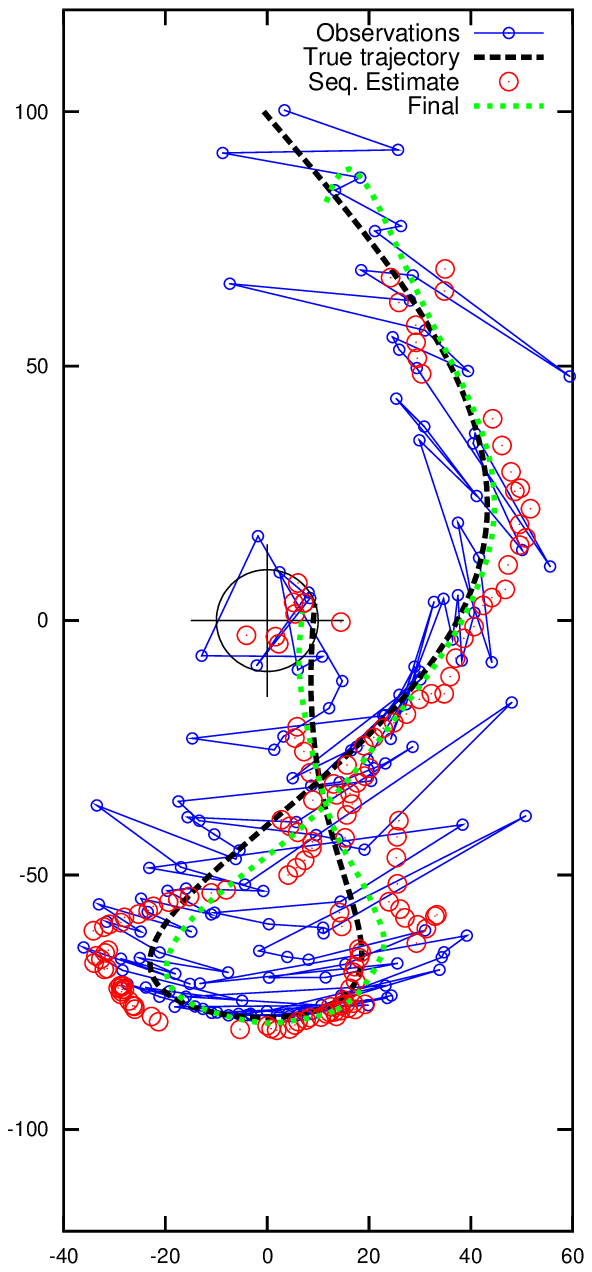}
  \end{tabular}
  \caption{Position tracking of same path as fig.~\ref{fig:p2d}. Circle
    with cross-hair is the observation site. Sequential position estimates
    and final trajectory estimate for $\eta=1000$.  (a)~Bearing
    measurement ten times more accurate than range. (b)~Range measurement
    ten times more accurate than bearing. In both (a) and (b) the
    estimation ignores correlations and treats each component as being
    independent. (c)~As case (b) but using correlation as computed from
    raw position estimate; this does worse than~(b), because the
    correlation is calculated relative to the raw position estimates,
    which are very misleading.}
  \label{fig:br}
\end{figure}

Figure~\ref{fig:br} shows the tracking of a target using range and bearing
information of different accuracy. In panels (a) and (b) the correlation
of the raw position estimates is ignored, and the results are good, that
is, the shadowing filter provides significant improvement over the raw
position estimates. Panel~(c) uses the same data as panel~(b), but tries
to take into account the correlation of the raw position estimates using
the correlation computed at the raw position estimate; the result is worse
than assuming no correlation. If the same is attempted for the data of
panel~(a), the result is a worse failure, because the radius is poorly
estimated and since this appears as a reciprocal in the transformation,
small errors in the radius can lead to very poorly estimated correlations,
so much so that the quality of the filtering is much worse. 

This example provides an excellent illustration of how the robustness of a
shadowing filter has significant gains over other filters. Not only does
ignoring the correlation result is better tracking, it is also more
efficient, because rather than using the vectorised
filter~(\ref{eq:masterd}), the simpler, more compact, scalar
filter~(\ref{eq:master}) can be used.

\subsection{Multiple bearing observations} Consider a target tracked in
the plane, position $\p=(x,y)$, using bearing observations
$q=(\theta,\theta')$ from two distinct reference points $(a,b)$ and
$(a',b')$. Under most circumstances there exists unique $s,s'\in\R$ such
that
\begin{eqnarray}
  \label{eq:br}
  (x,y) &=& (a,b) + s(\cos\theta,\sin\theta)\\
   &=& (a',b') + s'(\cos\theta',\sin\theta').
\end{eqnarray}
Hence, a raw position estimate can be found by solving the linear
equations
\begin{equation}\label{eq:lin}
  \begin{pmatrix}
    \cos\theta & -\cos\theta'\\
    \sin\theta & -\sin\theta'
  \end{pmatrix}
   \begin{pmatrix}
    s \\ s'
  \end{pmatrix}
  =
  \begin{pmatrix}
    a'-a \\ b'-b
  \end{pmatrix},
\end{equation}
provided the equations are consistent and non-singular. Since the angles
are observed angles, inconsistency can occur when
$\theta'\approx\pm\theta$. For some sequential filters such
nearly-singular situations can be devastating, but, since the shadowing
filter is estimating a trajectory from a sequence of observations, there
is generally no harm in simply dropping observations corrupted by
near-singularities, or replacing them with forecasted positions, or
crudely interpolated positions.  There is generally no harm in doing
either of these for short periods. Dropping observations requires using a
larger $\tau$ time-gap between observations. If the observations were
equally spaced in time, this leads to a lot of special computation, and so
in this case it is generally easier to insert a forecasted position with a
suitably scaled down information matrix to account for the errors in the
forecast; see details given later.

In this multiple-bearing situation it is difficult to compute~$J$
directly, but $K$ is easily computed. Note that
\begin{equation}
  \theta = \arctan\frac{y-b}{x-a} 
  \qquad\text{and}\qquad 
  \theta' = \arctan\frac{y-b'}{x-a'}. 
\end{equation}
It follows that
\begin{equation}
  K =
  \begin{pmatrix}
    (x-a)/r^2 & (y-b)/r^2\\
    (x-a')/r'^2 & (y-b')/r'^2
  \end{pmatrix},
\end{equation}
where $r^2=(x-a)^2+(y-b)^2$ and ${r'}^2=(x-a')^2+(y-b')^2$, which requires
only trivial computation once an estimate of $(x,y)$ is obtained.

\begin{figure}
  \centering
  \includegraphics[trim={90 30 90 30},width=\linewidth]{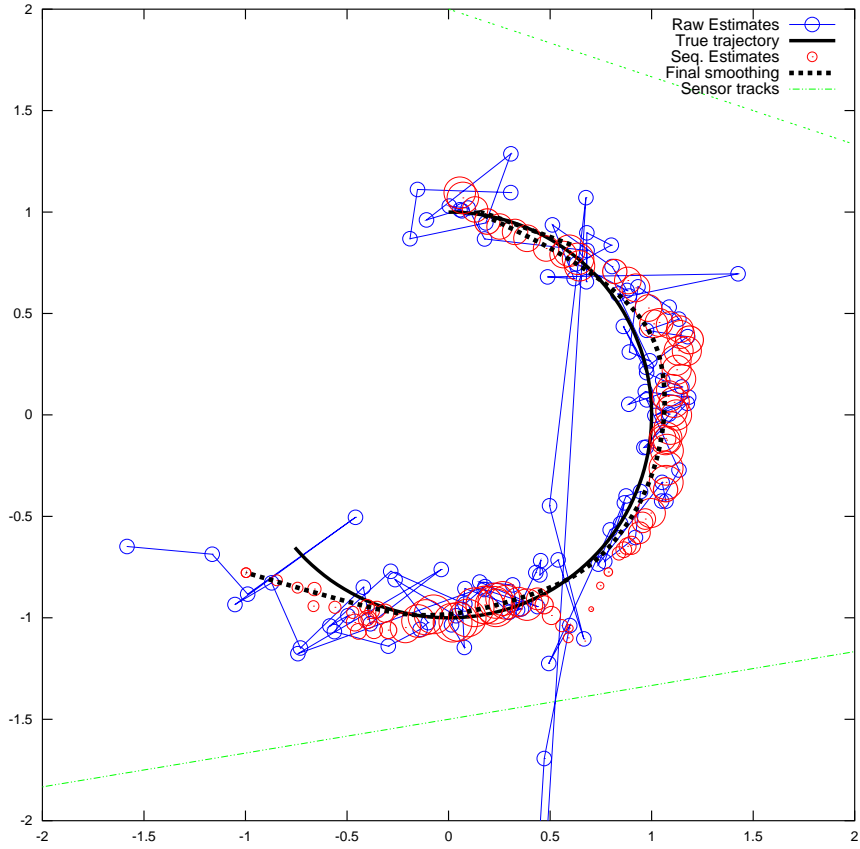}
  \caption{Position tracking using bearings from two moving sensors.  For
    $0\leq{t}\leq100$ one sensor moves between $(-3,3)$ and $(3,1)$, and
    the other between $(-3,-2)$ and $(3,-1)$, both at constant speed,
    while the target moves on the circular path
    $(\sin(t/25),\cos(t/25))$. The centre of the circles is the position
    of the sequential state estimates, the diameter of the circle indicates
    the condition-number weight applied to that position's raw position 
    estimate.}
  \label{fig:sonar}
\end{figure}

As mentioned, a significant problem arises when the sensors and target are
collinear, $\theta\approx\pm\theta'$, because the linear
equations~(\ref{eq:lin}) are singular or badly conditioned. This can
result in raw position estimates far their true
position. Figure~\ref{fig:sonar} shows an example of tracking using two
mobile sensors for detecting bearing, and a mobile target. In this example
the target moves on a circular path clockwise from the 12 o'clock
position. When the target is between the 4 and 5 o'clock position it is
directly between the sensors, and at the end of its path, around the 7
o'clock position, the target is almost directly behind both sensors. Both
of these situations lead to a poorly conditioned matrix in
eq.~(\ref{eq:lin}) and hence poor raw position estimates.

For the estimates shown in fig.~\ref{fig:sonar} the component correlations
of the raw position estimate are ignored, as in fig~\ref{fig:br}(a)
and~(b). The ill-conditioning is dealt with by using the 1-norm estimate
of the reciprocal condition as returned by LAPACK~\cite{LAPACK}. This
number varies between zero and one, with small values indicating bad
conditioning. This norm is a natural candidate to scale the information
matrices to give raw position estimates from poorly conditioned situations
small weight.

From fig.~\ref{fig:sonar} it can be seen that when the raw position
estimates are obtained under good conditioning, the sequential estimates
(large circles) are good. Under poor conditioning (small circles) the
sequential estimates are mainly forecasts from the preceding trajectory
positions, and the wild raw position estimates are ignored. When the
target passes beyond the 5 o'clock position and conditioning improves, and
the sequential estimates return to good position estimates. Note also that
the final trajectory almost exactly matches the true trajectory through
the 4 to 5 o'clock position.

\subsection{Multiple range observations} Consider a target tracked in the
plane, position $\p=(x,y)$, using range observations $q=(r,r')$ from two
distinct reference points $(a,b)$ and $(a',b')$. Under most circumstances
the location can be obtained from the solutions of $r^2=(x-a)^2+(y-b)^2$
and ${r'}^2=(x-a')^2+(y-b')^2$, assuming that the non-uniqueness can be
resolved. By taking partial derivatives of these two equations with respect
to $x$ and~$y$, it follows implicitly that
\begin{equation}
  K =
  \begin{pmatrix}
    (x-a)/r & (y-b)/r\\
    (x-a')/r' & (y-b')/r'
  \end{pmatrix}. 
\end{equation}

\section{Partial observations}
Our stated formulation of the tracking problem allows for non-uniformly
spaced observations, but all examples thus far have used only uniformly
spaced observations. We make a simple demonstration using non-uniformly
spaced observations by considering a situation where observations are
missing.

\begin{figure}
  \centering
  \includegraphics[width=\linewidth,trim=0 0 180 180]{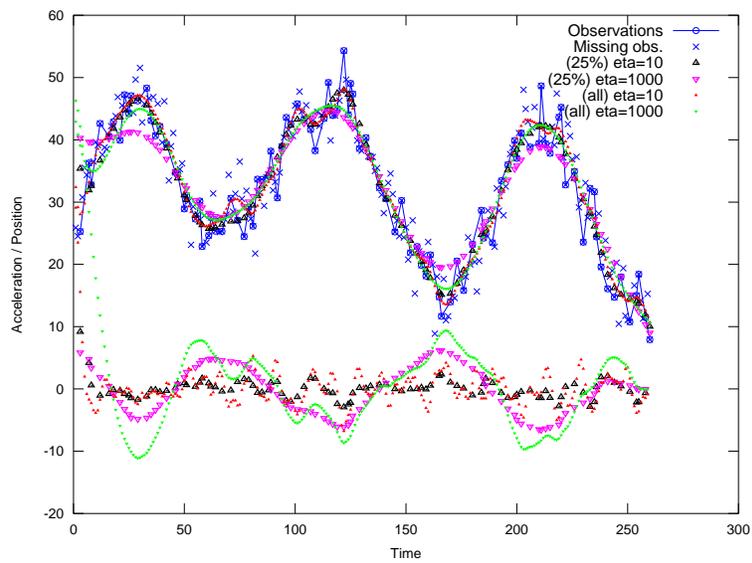}
  \caption{Position tracking and accelerations using similar data to
    figure \ref{fig:1a} and~\ref{fig:1b}, but with partial observations
    for two $\eta$ smoothing values. The first two curves show
    calculations when 75\% of observations are missing, with crosses
    marking the missing observations, and the second two curves using all
    observations.}
  \label{fig:pa2}
\end{figure}

Figure~\ref{fig:pa2} demonstrates tracking using similar data to
figure~\ref{fig:1a} where 75\% of the observations are missing, comparing
this to the tracking calculations when all observations are available. Two
values of the smoothing parameter~$\eta$ are used. These results
demonstrate that the tracking algorithm is very robust. The position
tracking is very similar when there is missing observations.  The
acceleration estimates are also very similar. Overall the tracking is
slightly smoother, and accelerations less variable, when there is missing
data, but this is something of an artifact, because the effective amount
of smoothing for a fixed $\eta$ depends on the amount of observations
available; less data results in more smoothing.

\section{Conclusion}\label{sec:notworthit}
Under the assumption of piecewise constant accelerations a shadowing
filter algorithm has been derived and implemented efficiently for scalar
time-series of observations. Vector time-series of observations can be
dealt with efficiently using the same algorithm if each component is
observed with uncorrelated errors.

The scalar algorithm can be easily extended to deal with tracking in
$d$-dimensions where the position vector components are not observed
directly and a transformation of the observations can be used to obtain an
initial \emph{raw position estimate}. The question then arises of how to
deal with the correlation of the errors this introduces. Remarkably,
experiments reveal, as illustrated in figs \ref{fig:p2d}, \ref{fig:br} and
\ref{fig:sonar}, that ignoring this correlation has little significant
effect. Simply using the raw position estimates to estimate the
correlations produced worse position estimates, because errors in the raw
position estimates give misleading indications about the correlation. This
problem is unvoidable to Kalman filters. It may be that some more complex
algorithm could be devised to better estimate the correlations, but this
will increase the amount of computation, for possibly no significant
gain. Just taking correlations into account in the stated algorithm
increases the size of matrix requiring singular value decomposition from
$n(n+1)$, to $d^2n(n+1)$ entries, so the computation cost is significant.

There are a number of other implementation issues that have not been
discussed, the most important of which is the optimal \emph{window}
size~$n$ for obtaining a shadowing trajectory and position estimates. The
window size is problem depended, but a method for determining an
appropriate window size is discussed at length
elsewhere~\cite{Stemler-Judd:guide1}. This cited work also discusses
issues of how best to implement sequential filtering.

\section*{Acknowledgements}
Supported by Australian Research Council Discovery Project DP0984659.

\bibliographystyle{plain}
\bibliography{refs}

\end{document}